\documentclass[aip,numerical,superscriptaddress,longbibliography]{revtex4-2}
\usepackage{graphicx}
\usepackage{epstopdf}
\usepackage{epsfig}
\usepackage{mathrsfs}
\usepackage{amssymb}
\usepackage{upgreek}
\usepackage{bbold}
\usepackage{color}
\usepackage{bm}
\usepackage{longtable}

\begin{document}
%
%
%
%
\title{Generalization of multivariable Laplace transform based on Tsallis $q$-exponential and its inverse using Post-Widder's method}

\author{Naina Mohammed Samu Shahabuddin}
\email[]{naina@gacudpt.in}
\affiliation{Department  of Physics, Government Arts College, Udumalpet, Tiruppur District  - 642126, Tamilnadu, India}

\author{Jeevanandham Karthikeyan}
\affiliation{Department  of Physics, Government Arts College, Udumalpet, Tiruppur District  - 642126, Tamilnadu, India}

\author{Basherrudin Mahmud Ahmed Abduljaffer}
\affiliation{School of Physics, Madurai Kamaraj University, Madurai 625021, India}

\author{Md.~Manirul Ali}
\affiliation{Centre for Quantum Science and Technology, Chennai Institute of Technology, Chennai 600069, India}

\author{Chandrashekar Radhakrishnan}
\affiliation{Department of Computer Science, Data Science and Engineering, NYU Shanghai, 567 Yangsi West Road, Pudong New District, Shanghai 200124, China}
\affiliation{Centre for Quantum Information, Communication and Computing, Indian Institute of Technology Madras, Chennai 600036, India}

\begin{abstract}
A generalization of the multivariate Laplace transform based on the Tsallis $q$-exponential is given in the present work for a new type of kernel.  
We also define the inverse transform for this generalized multivariate transform based on the complex integration method. 
We prove identities corresponding to the Laplace transform and inverse transform like the $q$-convolution theorem, the action of generalized 
derivative and generalized integration on the Laplace transform.  We then derive a $q$-generalization of the inverse Laplace transform 
based on the Post-Widder's method which bypasses the necessity for a complex contour integration.  This extension of the Post-Widder's 
method has been carried out for the multivariate case as well.  We demonstrate the usefulness of this in computing the Laplace and inverse 
Laplace transform of some elementary functions.  The inverse Laplace transform has been calculated using both the complex integration 
method as well as the Post-Widder's methods and the results obtained through these two methods agree with each other.  
\end{abstract}

\maketitle


%
%
%
\section{Introduction}
\label{Intro}
A mathematical function `$f$' in the `$x$' domain can be transformed to a function `$F$' in the `$u$' domain 
using the integral 
\begin{equation}
F(u) = \int_{a}^{b} f(x)  \, K(u,x) dx.
\label{ltone_def}
\end{equation}
This process is known as an integral transform and the function $K(u,x)$ is the kernel of the transformation. 
An extension of Eq. (\ref{ltone_def}) to the multivariate case reads: 
\begin{equation}
F(u_{1},...,u_{n}) = \int_{a_{1}}^{b_{1}} \cdots \int_{a_{n}}^{b_{n}} f(x_{1}, ... , x_{n}) K(u_{1},...,u_{n},x_{1},...,x_{n}) 
dx_{1} ... dx_{n}.
\end{equation}
Here the integral transform maps the problem in a given domain to another domain in which it is simpler to 
solve.  Laplace transform is one of the widely used integral transforms in physics.  It is very useful in solving
convolution integral equations and differential equations.  The inverse Laplace transform is an equally 
important transform with several applications.  In general an inverse Laplace transform is carried out using 
a Bromwich contour integral in the complex plane  \cite{schiff1999laplace} .  But this practice of using a complex 
variable technique is done only for the sake of convenience.  In fact an inverse Laplace transform based only on real 
variables was introduced by Post \cite{post1930generalized} and later it was refined by Widder \cite{widder1934inversion}.  
This method has been investigated in several works \cite{jagerman1982inversion,al2001inversion,soldatov2000widder}
 considering different applications.  

The kernel of a Laplace transform is an exponential function of the form $\exp(-st)$.  Tsallis introduced 
\cite{tsallis1988possible} generalizations of the logarithm and the exponential functions as follows: 
\begin{equation}
\ln_{q}(x) \equiv  \frac{x^{1-q} - 1} {1-q};  \qquad  \exp_{q}(x) \equiv [1 + (1-q) x]^{\frac{1}{1-q}},
\end{equation}
where $q \in \mathbb{R}_{+}$ is the generalization parameter.  These functions are generally referred to 
as Tsallis $q$-logarithm and Tsallis $q$-exponential and are inversely related.  The $q$-logarithm 
and the $q$-exponential functions reduce to the usual logarithm and exponential functions in 
the $q \rightarrow 1$ limit.  These generalized functions have been investigated in a wide variety 
of fields like astrophysics \cite{nakamichi2002non,sakagami2004self}, high energy physics \cite{tsallis2003nonextensive,tsallis2003fluxes},
 neutrino physics \cite{luciano2021nonextensive}, mathematical physics \cite{umarov2022mathematical,yamano2002some} and 
nonequilibrium statistical physics \cite{rajagopal1998equations}.  

A generalization of the Laplace transform has been done in Ref. \cite{lenzi1999q,plastino2013tsallis,chung2013q,naik2016q} 
using the Tsallis q-exponential.  So far only the single variable Laplace transform has been considered in these works.  
In our present work we study the multivariable Laplace transform based on Tsallis $q$-exponential.  Further the inverse transforms
has not been defined for even the single variable Laplace transform introduced in \cite{naik2016q}.  First we define the inverse using 
a contour integral in the complex plane for the Laplace transform defined in \cite{naik2016q}.  Then we use the Post-Widder's 
method  \cite{post1930generalized,widder1934inversion} to compute the inverse transform using real variable.  
In our work we introduce the inverse for the Laplace transform based on the generalized exponential function.  We also 
extend this generalized Laplace transform and its inverse to the case of multivariable functions 
\cite{coon1953some,lubbock1969multidimensional,jensen1997convergence,eltoft2006multivariate} .

The work is organized as follows:  In Section 2, we give a brief introduction to the Post-Widder's technique and the Tsallis $q$-exponential.  
The single and multivariable Laplace transform based on type - I kernel is defined in the third section.  The inverse transform based on 
complex integration method is also defined in the same section.  The properties of Laplace transform are derived in Section 4. In Section 5, we 
give the $q$-generalization of the Post-Widder's method for computing the inverse Laplace transform for both the single variable and the 
multivariable case.  The Laplace transform and the inverse Laplace transform based on the Widder's method for a simple function is computed in 
Section 6.  We also compile a table consisting of the $q$-generalized Laplace transform and its inverse for some elementary functions.  For the 
inverse transform the calculations were done using both the complex integration method and the Post-Widder's method and the results obtained
through these two different methods agree with each other.

%
%
%

\section{A primer on Laplace transform, Post- Widder's technique and Tsallis $q$-exponential}
\label{primer}
In this section we briefly review the concept of Laplace transform and its inverse.  Then we describe the 
Post-Widder's method to calculate the inverse Laplace transform.  The Laplace transform of a function 
$f(t)$ denoted by $\mathcal{L}[f(t)] = F(s)$ is defined as 
\begin{equation}
F(s) = \int_{0}^{\infty} \exp(-st) f(t) dt. 
\label{laplacetransform}
\end{equation}
Here $f \in S(\mathbb{R})$ where $S(\mathbb{R})$ represents the Schwartz space of functions 
$f: \mathbb{R} \rightarrow \mathbb{C}$ with $f \in C^{\infty} (\mathbb{R})$, i.e., $f$ is infinitely differentiable
on $\mathbb{R}$.  In general $s = \sigma + i \tau$ with $\sigma$ and $\tau$ being real numbers.  The 
integral converges when ${\rm Re}[s] = \sigma > 0$ and for $\sigma < 0$, $F(s) =0$.  
For the Laplace transform of $F(s)$, the inverse transform is defined as 
\begin{equation}
f(t) = \frac{1}{2 \pi i}   \lim_{\tau \rightarrow \infty}  \int_{\sigma - i \tau}^{\sigma+i \tau}  \exp(st) F(s) ds.
\label{inverselaplacetransform}
\end{equation}
The direct Laplace transform Eq. (\ref{laplacetransform})  and the inverse Laplace transform Eq. (\ref{inverselaplacetransform}) 
are inverses of each other for the functions $f \in S(\mathbb{R})$.   In order to calculate the inverse Laplace transform we need to perform a 
Bromwich contour integration over the complex plane. Below we explain an alternative method introduced 
by Post and Widder which does a Laplace inverse using only real variables.  

Let us consider the $n^{th}$ derivative of the Laplace transform $F(s)$ with respect to the variable `$s$', 
\begin{equation}
\frac{ {\rm d}^{n}} {{\rm d} s^{n}} F(s)  \equiv F^{(n)}(s) = (-1)^{n} \int_{0}^{\infty} t^{n} \exp(-st) f(t) dt.
\label{laplacederiv}
\end{equation}
We use the following three steps,  (i) First we use the variable transformation $s =n/x$,  
(ii) We follow it by multiplying the numerator and the denominator by $x^{n}$  and  (iii) Finally
we use the variable change $y= t/x$, and recast the integral in Eq. (\ref{laplacederiv}) to 
\begin{equation}
F^{(n)} \left( \frac{n}{x} \right) = (-1)^{n}  x^{n+1} \int_{0}^{\infty}  (y e^{-y})^{n} f(xy) dy.
\label{laplaceintegral}
\end{equation}
Here we would like to point out that the function $ y e^{-y}$ has a single maximum at $y=1$ and for 
the function $(y e^{-y})^{n}$ this maximum is sharply peaked at $y=1$ and hence Eq. (\ref{laplaceintegral}) can be 
rewritten as
\begin{equation}
F^{(n)} \left( \frac{n}{x} \right)  \approx (-1)^{n} x^{n+1} f(x) \int_{0}^{\infty} (y e^{-y})^{n} dy.
\end{equation}
Evaluating the integral we get
\begin{equation}
F^{(n)} \left( \frac{n}{x} \right)  \approx (-1)^{n}  n! \left( \frac{x}{n} \right)^{n+1} f(x).
\label{laplaceintresult}
\end{equation}
In  the limit $n\rightarrow \infty$, we can use the transformation 
$y=\frac{t}{x} \big|_{y=1} \Rightarrow x=t$ and so $s=\frac{n}{x} \Rightarrow s=\frac{n}{t}$ we can rewrite 
Eq. (\ref{laplaceintresult}) as 
\begin{equation}
f(t) = \lim_{n \to \infty}  \frac{(-1)^{n}}{n!} \, s^{n+1} F^{(n)} (s)  |_{s=n/t}.
\end{equation}
Thus we observe that the Laplace transform and the inverse Laplace transform can be expressed 
as functions of real variables alone.  This sequence converges very fast since the rate of convergence 
is at least $1/n$.  

The Laplace transform exists for a piecewise continuous function of exponential order. Similarly, the 
$q$-Laplace transform has been defined for a piecewise continuous function of q-exponential order.  
The $q$-exponential based  Laplace transform can be defined using three different types of kernels 
as noted in Ref. \cite{lenzi1999q} and these kernels are 
\begin{eqnarray}
K_{{\rm I}} (q;s,t) &=& \exp_{q}(-st),  \\
K_{{\rm II}}(q;s,t) &=& [\exp_{q}(-t)]^{s}, \\
K_{{\rm III}}(q;s,t) &=& [\exp_{q}(t)]^{-s}. 
\end{eqnarray}
Of these three kernels, Laplace transform has been defined using the first and the second kernels 
in Ref .  \cite{lenzi1999q,chung2013q,naik2016q}.
In the extensive limit ($q \rightarrow 1$), both these generalized Laplace transforms reduce to the 
ordinary Laplace transform.  For the generalized Laplace transform based on the first kernel \cite{naik2016q}, the 
inverse Laplace transform has not been defined.  But for the case where the Laplace transform was 
defined using the second kernel, the inverse transform was defined using the complex integration method.

%
%
%

\section{Laplace and Inverse transform based on type - I kernel}
\label{typeIkernel}
The Laplace transform due to type-I kernel was introduced in \cite{naik2016q} for a single variable.  But an 
introduction of the inverse transform has not been done so far.  In the present section we have two subsections, 
where the first one discusses the Laplace and inverse Laplace transform of single variable function.  The second 
subsection introduces the multivariable Laplace transform and its inverse.  Throughout this section we restrict 
and present the results only  for the $q < 1$ case.  

\subsection{Single variable Laplace transform:} 
The single variable Laplace transform introduced in Ref. \cite{naik2016q} is revisited here.  However an inverse 
Laplace transform has not been defined so far.  Here in our work we define the inverse Laplace transform and also 
prove its inverse property.  
A function `$f$' is said to be of $q$-exponential order `$c$', if there exists $c$, $M >0$, $T>0$, such that 
$|f(t)| \leq M \exp_{q}(ct)$  $\forall$ $t>T$.  If a function is piecewise continuous and is of $q$-exponential order 
$c$, then $F_{q}(s) = L_{q} [f(t)]$ exists for $s>c$ and $\lim_{s \rightarrow \infty}  F(s) =0$. 
Under these conditions, the $q$-Laplace transform is defined as 
\begin{equation}
L_{q} [f(t)](s) = F_{q}(s) = \int_{0}^{\infty} f(t) \exp_{q}(-st) dt.
\label{laplacetypeI}
\end{equation}
The corresponding inverse Laplace transform is defined as 
\begin{equation}
L_{q}^{-1} [ F_{q}(s) ] (t) = f(t) = \frac{2-q}{2 \pi i}  \int_{c-i \infty}^{c+i\infty} F_{q}(s)  [\exp_{q}(-st)]^{2q-3}  ds,
\end{equation}
where in the limit $q \rightarrow 1$, we have $[\exp_{q}(-st)]^{2q-3} \rightarrow \exp(st) $.
Here $c$ is a real constant that exceeds real part of all the singularities of $F_{q}(s)$.  To prove the inverse 
relationship between $L_{q}$ and $L_{q}^{-1}$, we verify the following two identities:
\begin{eqnarray}
f(t) &=& L_{q}^{-1} [ L_{q}  [ f(t) ] ],
\label{Identity1}  \\
F_{q}(s) &=& L_{q} [ L_{q}^{-1} [ F_{q} (s) ] ]. 
\label{Identity2}
\end{eqnarray}
Using the inverse Laplace transform we can write (\ref{Identity2}), 
\begin{eqnarray}
 L_{q} [ L_{q}^{-1} [ F_{q} (s) ] ]  &=&  \int_{0}^{\infty}  \exp_{q}(-st) \; L_{q}^{-1} [ F_{q} (s) ]  dt  \nonumber  \\
&=&  \int_{0}^{\infty} dt  \exp_{q}(-st) \; \frac{(2-q)}{2 \pi i } \;  
         \int_{c-i \infty}^{c+i \infty}  ds^{\prime}  \; F_{q}(s^{\prime})  [ \exp_{q}(-s^{\prime} t) ]^{2q-3}.
\end{eqnarray}
We can rewrite the above expression as
\begin{equation}
 L_{q} [ L_{q}^{-1} [ F_{q} (s) ] ]  =  \frac{(2-q)}{2 \pi i } \;  \int_{c-i \infty}^{c+i \infty}  ds^{\prime}  \; F_{q}(s^{\prime})  
                                                          \int_{0}^{\infty} dt  \exp_{q}(-st) \; [ \exp_{q}(-s^{\prime} t) ]^{2q-3}.
\label{laplacelaplaceinverse}                                                          
\end{equation}
The second integral converges when ${\rm Re}[s^{\prime}] = c < {\rm Re}[s]$ and the resulting solution is 
\begin{equation}
\mathcal{I}_{q} (s,s^{\prime}) =  \int_{0}^{\infty} dt  \exp_{q}(-st) \; [ \exp_{q}(-s^{\prime} t) ]^{2q-3} = \frac{1}{(2-q)} \;  \frac{1}{(s-s^{\prime})}. 
\end{equation}
Substituting this in Eq. (\ref{laplacelaplaceinverse}) we get
\begin{equation}
L_{q} [ L_{q}^{-1} [ F_{q} (s) ] ]  = \frac{1}{2 \pi i }  \;  \int_{c-i \infty}^{c+i \infty} ds^{\prime} \frac{F_{q}(s^{\prime})} {(s-s^{\prime})}.
\label{laplapinverse}
\end{equation}
The integrand $F_{q}(s^{\prime})/(s-s^{\prime})$ has a simple pole at $s=s^{\prime}$.  To evaluate the integral we draw a 
straight line at $s^{\prime}$ and an arc enclosing the pole of the integrand.  The solution of the integral is 
\begin{equation}
\int_{c-i \infty}^{c+i \infty} ds^{\prime} \frac{F_{q}(s^{\prime})} {(s-s^{\prime})} = 2 \pi i  F_{q} (s). 
\end{equation}
Substituting this in Eq. (\ref{laplapinverse}) we can observe that $L_{q} [ L_{q}^{-1} [ F_{q} (s) ] ] = F_{q} (s)$.  
Next we verify the first identity Eq. (\ref{Identity1}) as follows:
\begin{equation}
L_{q}^{-1} [ L_{q} [ f(t) ] ]  \equiv L_{q}^{-1} [F_{q}(s)] =  L_{q}^{-1} \left[  \int_{0}^{\infty}  dt \exp_{q}(- st) f(t) \right]. 
\end{equation}
Using the definition of the inverse Laplace transform 
\begin{eqnarray}
L_{q}^{-1} [ L_{q} [ f(t) ] ] = \frac{2-q}{2 \pi i} \int_{c-i \infty}^{c+\infty}  [ \exp_{q}(-s t) ]^{2q-3}  
                                     \int_{0}^{\infty}  dt^{\prime} f(t^{\prime}) \exp_{q}(-s t^{\prime}) \;  ds. 
\label{lapinverselap}                                             
\end{eqnarray}
Rewriting the above equation (\ref{lapinverselap}) we arrive at 
\begin{equation}
L_{q}^{-1} [ L_{q} [ f(t) ] ] = \frac{2-q}{2 \pi i} \;  \int_{0}^{\infty}  dt^{\prime} f(t^{\prime})  
                                             \int_{c-i \infty}^{c+\infty}  [ \exp_{q}(-s t) ]^{2q-3}   
                                        \exp_{q}(-s t^{\prime})  \;  ds.                                    
\end{equation}
Using the definition of the Dirac delta function based on the $q$-exponential described in 
Ref.  \cite{mamode2010integral,jauregui2010new} we get
\begin{equation}
L_{q}^{-1} [ L_{q} [ f(t) ] ] =  \int_{0}^{\infty}  dt^{\prime} f(t^{\prime})  \delta(t-t^{\prime}) \equiv f(t). 
\end{equation}
Thus in the present section,  we have introduced the inverse for the Laplace transform defined using the 
type-I kernel using the complex integration technique.  

\subsection{Multivariable Laplace transform:} 
A generalization of the $q$-Laplace transform to the case of a multivariable function is carried out in this section. 
Let us consider a multivariable function $f(t_{1},..., t_{n})$ for which the $q$-generalization of the multivariable 
Laplace transform is defined as follows: 
\begin{eqnarray}
L_{q}\left[f(t_{1}, \ldots,t_{n})\right] &=& \displaystyle \int_{0}^{\infty} d t_{1},  
							     \ldots \int_{0}^{\infty} dt_{n} \;
							     \exp_{q} \left( - \sum_{l=1}^{n} s_{l} t_{l} \right) f(t_{1}, \ldots,t_{n}) \nonumber \\
						      &=&   F_{q} (s_{1}, \ldots, s_{n})
\label{multiLT}						      
\end{eqnarray}
In the limit $q \rightarrow 1$, the generalized Laplace transform defined in Eq. (\ref{multiLT}) reduces to the ordinary 
Laplace transform.  The inverse transform corresponding to the multivariable Laplace transform is 
\begin{eqnarray}
f(t_{1}, \ldots, t_{n}) &=& \frac{Q_{n}(2-q)}{(2\pi i)^{n}} \displaystyle 
					     \int^{c_{1} + i \infty}_{c_{1} - i \infty} ds_{1} 
					    \ldots \int^{c_{n} + i \infty}_{c_{n} - i \infty} ds_{n}  
                         \left[  \exp_{q} \left(	- \sum_{l=1}^{n} s_{l} t_{l} \right) \right]^{\frac{(n+1)q - (n+2)}{1-q}}   \nonumber \\
                        & & \times  F_{q} (s_{1}, \ldots, s_{n})		     
\end{eqnarray}
where $c_{1},...,c_{n}$ are real constants that exceeds real parts of the singularities and the factor
$Q_{m+1}(2-q) = (1-q)^{m+1} \frac{\Gamma\left(\frac{1}{1-q}+m+2\right)}{\Gamma\left(\frac{1}{1-q} + 1\right)}$.

%
%
%
\section{Properties of $q$-Laplace transform}
\label{properties}
In this section, we list some of the properties of the $q$-Laplace transform: 
\begin{enumerate}
    \item ${\rm I}^{st}$ Identity on Limits: 
    \begin{equation}
        \lim_{s \rightarrow \infty} s L_{q}[f(t)] = 
        \lim_{t \rightarrow 0}  \frac{f(t)}{1+(1-q)}, 
    \end{equation}  
    {\it Proof:}  
    Let us consider a general convergent function which can be expressed in terms of a power series 
    $f(t) = \sum_{n=0}^{\infty} a_{n} t^{n}$.  The $q$-Laplace transform of this general function is 
    \begin{equation}
        L_{q} [f(t)] = \int_{0}^{\infty} \left(\sum_{n=0}^{\infty} a_{n} t^{n} \right) [1-(1-q)st]^{\frac{1}{1-q}} \, dt .
        \label{lc11}
    \end{equation}
    Since the function $f(t)$ is a convergent function we can rewrite the above Equation (\ref{lc11}) as 
     \begin{equation}
        L_{q} [f(t)] =   \sum_{n=0}^{\infty} a_{n} \int_{0}^{\infty} t^{n} [1-(1-q)st]^{\frac{1}{1-q}} dt .
        \label{lc12}
    \end{equation}
    To solve Eq. (\ref{lc12}) we use the integration by parts method choosing $u = t^{n}$ and 
    $dV = [1-(1-q)st]^{\frac{1}{1-q}}$.  The resulting solution reads: 
    \begin{eqnarray}
       s L_{q} [f(t)]  &=&  -  \frac{1}{1+(1-q)} \bigg( f(t) [1- (1-q) st]^{\frac{1}{1-q} + 1}  \big|_{0}^{\infty}  \nonumber \\
                       & &  - \int_{0}^{\infty} [1-(1-q) st] \left [ \frac{d}{dt} f(t) \right]  \exp_{q} (-st) dt \bigg) .
                       \label{lc13}
    \end{eqnarray}
    On applying the limits corresponding to the integration in Eq. (\ref{lc13}) we get 
    \begin{eqnarray}
       s L_{q} [f(t)] &=& \frac{f(0)}{1+(1-q)} + \frac{1}{1+(1-q)} 
       \nonumber \\
       & &  \int_{0}^{\infty} [1-(1-q)st] \; \frac{d f(t)}{dt} 
                         [1 - (1-q) st]^{\frac{1}{1-q}} dt.
                         \label{lc14}
    \end{eqnarray}
    Under the limiting condition $s \rightarrow \infty$, Eq. (\ref{lc14}) gives 
    \begin{equation}
        \lim_{s \rightarrow \infty} s L_{q}[f(t)] =
        \lim_{s \rightarrow \infty}  \frac{f(0)}{1+(1-q)} .
    \end{equation}
   We can observe that the RHS is independent of $s$ and can be expressed as a limiting value of the parameter $t$ and this gives us 
    \begin{equation}
        \lim_{s \rightarrow \infty} s L_{q}[f(t)] =
        \lim_{t \rightarrow 0}  \frac{f(t)}{1+(1-q)}.
    \end{equation}
  Thus we prove the first identity on the limits of a Laplace transform. 

  \item ${\rm II}^{nd}$ Identity on Limits: 
    \begin{equation}
        \lim_{s \rightarrow 0} s L_{q}[f(t)] = 
        \lim_{t \rightarrow \infty}  \frac{f(t)}{1+(1-q)}.
    \end{equation}
    
    {\it Proof:} 
    To prove this limit let us consider the Eq. (\ref{lc13}) and evaluate the limit, $s \rightarrow 0$ and 
    the resulting expression is 
    \begin{eqnarray}
       \lim_{s \rightarrow 0} s L_{q} [f(t)] &=&  \lim_{s \rightarrow 0} \frac{f(0)}{1+(1-q)} 
                                + \frac{1}{1+(1-q)} \int_{0}^{\infty} \frac{df}{dt} dt \nonumber \\
                                &=& \frac{f(0)}{1+(1-q)} + \frac{f(t)}{1+(1-q)} \bigg|_{0}^{\infty}.
    \end{eqnarray}
    Applying the limits we get 
    \begin{equation}
       \lim_{s \rightarrow 0} s L_{q} [f(t)] =  \frac{f(\infty)}{1+(1-q)}.
    \end{equation}
    The RHS of the equation can be rewritten as 
    \begin{equation}
        \lim_{s \rightarrow 0} s L_{q} [f(t)] = \lim_{t  \rightarrow \infty}  \frac{f(t)}{1+(1-q)}.
    \end{equation}
    Thus we prove the second identity on the Laplace transform.  
    
    \item Scaling:
    \begin{equation}
        L_{q} [ f(at) ] = \frac{1}{a} F_{q} (s/a).
    \end{equation}
    
    {\it Proof:} Let us consider the Laplace transform of a function $f(at)$ 
    \begin{equation}
        L_{q} [ f(at) ] = \int_{0}^{\infty} dt [1-(1-q) st]^{\frac{1}{1-q}} f(at).
    \end{equation}
    Substituting $at = x$, we get 
    \begin{equation}
        L_{q} [ f(at) ] = \frac{1}{a} \, \int_{0}^{\infty} dx  \bigg[1-(1-q) \frac{sx}{a} \bigg]^{\frac{1}{1-q}} f(x) 
                          = \frac{1}{a} F_{q}(s/a).
    \end{equation}
    Hence the scaling relation is proved. 
    
    \item Shifting:
    \begin{equation}
        F_{q}(s-s_{0}) = L_{q} \left[ f(t) \exp_{q} \left[ \frac{s_{0} t}{1-(1-q) s t} \right] \right] .
    \end{equation}
    {\it Proof:} Let us consider the definition of the Laplace transform 
    \begin{equation}
        F_{q}(s) = \int_{0}^{\infty} \exp_{q}(-st) f(t) dt,
    \end{equation}
    and introduce a shift in `$s$' as `$s \rightarrow s-s_{0}$' and this yields 
    \begin{equation}
        F_{q}(s-s_{0}) = \int_{0}^{\infty} \exp_{q}(-(s-s_{0})t) f(t) dt.
    \end{equation}
    This can be rewritten as 
    \begin{equation}
        F_{q}(s-s_{0}) = \int_{0}^{\infty} \exp_{q}(-st) \exp_{q} \bigg( \frac{s_{0} t}{1-(1-q)st} \bigg) f(t) dt.
    \end{equation}
    Hence we have proved
    \begin{equation}
        F_{q}(s-s_{0}) = L_{q} \bigg[ f(t) \exp_{q} \bigg( \frac{s_{0} t}{1-(1-q)st} \bigg) \bigg]. 
    \end{equation}
    
    \item $q$-translation:
    \begin{equation}
        L_{q} \left [ f(t) \big[ \exp_{q}(st_{0}) \big]^{1+(1-q)}  \right] = 
             L_{q} \bigg[ f\left(\frac{t-t_{0}}{1 - (1-q) s t_{0}} \right) \Theta \left(  \frac{t - t_{0}}{1-(1-q)s t_{0}} \right) \bigg].
    \end{equation}
    
    {\it Proof:}
    The RHS of the above identity gives 
    \begin{eqnarray}
    I &\equiv& L_{q} \bigg[ f\left(\frac{t^{\prime}-t_{0}}{1 - (1-q) s t_{0}} \right) \Theta \left(  \frac{t^{\prime} - t_{0}}{1-(1-q)s t_{0}} \right) \bigg] \nonumber \\
    &=& \int dt^{\prime} f\left(\frac{t^{\prime}-t_{0}}{1 - (1-q) s t_{0}} \right) 
    \Theta \left(  \frac{t^{\prime} - t_{0}}{1-(1-q)s t_{0}} \right) 
       \exp_{q}(-s t^{\prime}), \nonumber
    \label{qtrans1}      
    \end{eqnarray}  
    where $\Theta(x)$ is the Heaviside step function such that $\Theta(x) = 0$ when $x<0$ and $\Theta(x) = 1$ for $x \geq 0$.
    Using the scaling $t = \frac{t^{\prime} - t_{0}}{1-(1-q)s t_{0}}$, we can rewrite the integral as 
    \begin{equation}
        I = \int_{- \alpha}^{\infty} \; dt \exp_{q}(-st) \; 
        f(t) [\exp_{q}(-s t_{0})]^{2-q} \; \Theta(t).
    \end{equation}
    where $\alpha = \frac{t_{0}}{1-(1-q)s t_{0}}$.  On applying the Heaviside step function we get
    \begin{equation}
        I = L_{q}[f(t) (\exp_{q}(-s t_{0}))^{2-q}] .
    \end{equation}
    Hence the $q$-translation identity has been proved.  
\end{enumerate}

The properties of linearity and $q$-convolution have been established in Ref. \cite{naik2016q} and here we are stating 
them just for the sake of completeness.  The two properties of linearity and $q$-convolution are 
\begin{enumerate}
    \item Linearity: 
    \begin{equation}
     L_{q} [a_{1} f_{1}(t) + a_{2} f_{2}(t) ]  =
     a_{1} L_{q} [ f_{1}(t) ] +  a_{2} L_{q} [ f_{2}(t) ]. 
    \end{equation}
    
    \item $q$-convolution:  \\
    Let $f(t)$ and $g(t)$ be two positive scalar functions of `$t$', and $F_{q}(s)$ and $G_{q}(s)$ be 
    their q-Laplace transforms, then 
    \begin{equation}
        L_{q}[f(t) \ast g(t)] = F_{q}(s) \ast G_{q} (s), 
    \end{equation}
    where $f(t) \ast g(t)  = \int_{0}^{t} f(\tau) \ast g(t-\tau) d \tau$.
\end{enumerate}

\noindent{\it Laplace transform of $q$-derivatives and $q$-integrals:}
The properties of the $q$-Laplace transform based on derivatives and integrals have been derived in Ref. \cite{naik2016q}. 
For the sake of completeness, we give below the expression for the generalized Laplace transform of a derivative function
\begin{eqnarray}
 L_{q} \left[ \frac{{\rm d}^{n}} {{\rm d} t^{n} } f(t) \right] &=& 
- \left(f^{(n-1)}(t) + (1-\delta_{1n}) \sum_{\ell=1}^{n-1} Q_{\ell -1} (q) s^{\ell} f^{(n-\ell-1)}(t)  \right) \Bigg\vert_{t=0} \nonumber \\
   & & + Q_{n-1}(q) s^{n} L_{\frac{a_{n+1}}{a_{n}}} (f(t)) (a_{n} s),
\end{eqnarray}
where, $Q_{n}(q) = \prod_{j=0}^{n} a_{j}$ with $a_{j} = jq - (j-1)$ and 
$\delta_{1j}$ is the Kronecker delta function. The Laplace transform of the integral reads: 
\begin{equation}
    L_{q} \left[ \int_{0}^{t}  f(x) dx \right] = \frac{2-q}{s} L_{\frac{1}{2-q}}  \left[ f(t) \right] \left(s(2-q)\right).
    \label{integrallaplacetransform}
\end{equation}
In the present work, we derive the $q$-Laplace transform of $q$-calculus of functions.  The concept of 
$q$-calculus was first introduced in \cite{borges2004possible}, where the authors defined the derivative and integral based on 
$q$-deformation.  Many such derivatives and integrals were investigated in subsequent works \cite{umarov2022mathematical,lenzi1999q,naik2016q}. For the 
present work we use the $q$-derivative defined in \cite{chakrabarti2010nonextensive} and its corresponding $q$-deformed integral.  The
relevant $q$-derivative and $q$-integral operators are 
\begin{equation}
  D_{q}(s) = \frac{1}{1-(1-q) \left( s \frac{d}{ds} \right)}  \frac{d}{ds};  \qquad \int d_{q} x = \int dx \left[1-(1-q) x \frac{d}{dx}\right].
\end{equation}
\begin{enumerate}
    \item Derivative of Laplace transform:  
     \begin{equation}
         D_{q}^{(n)}(s) \{ L_{q} [f(t)] \} (s) = L_{q}[(-t)^{n} f(t)].
     \end{equation}

\noindent{\it Proof:} Let us consider the first derivative of the $q$-Laplace transform
\begin{eqnarray}
    D_{q} \{ L_{q} [f(t)] \} (s) &=& D_{q}(s) \left \{ \int_{0}^{\infty} dt \exp_{q}(-st) f(t)     \right \} \nonumber \\
                                 &=& \int_{0}^{\infty} dt \,  D_{q}(s) \exp_{q}(-st) f(t) \nonumber \\
                                 &=& L_{q}[(-t) f(t)].
\end{eqnarray} 
The second derivative of the $q$-Laplace transform is 
\begin{equation}
    D_{q}^{(2)} \{ L_{q} [f(t)] \} (s) = L_{q}[(-t)^{2} f(t)].
\end{equation}
For the $n^{th}$ $q$-derivative we get 
\begin{equation}
    D_{q}^{(n)} \{ L_{q} [f(t)] \} (s) = L_{q}[(-t)^{n} f(t)].
\end{equation}

\item Integral of a Laplace transform: 
\begin{equation}
    \int_{s}^{\infty} d_{q}s^{\prime} \, F_{q}(s^{\prime})  =  L_{q} \left\{ \frac{f(t)}{t} \right\}.
\end{equation}
Let us consider the definition of the generalized Laplace transform apply the $q$-integral 
operator
\begin{equation}
    \int_{s}^{\infty} d_{q} s^{\prime} \, F_{q}(s^{\prime}) = \int_{s}^{\infty} d_{q} s^{\prime} \int_{0}^{\infty} \exp_{q}(-s^{\prime} t) f(t) dt. 
\end{equation}
Rearranging the order of the integrals and evaluating the $q$-integral 
with respect to `$s$' gives
\begin{equation}
    \int_{s}^{\infty} d_{q} s^{\prime} \, F_{q}(s^{\prime}) = \int_{0}^{\infty} \frac{1}{t} \exp_{q}(-st) f(t) dt = L_{q} \left[ \frac{f(t)}{t} \right].
\end{equation}
Hence proved. 
\end{enumerate}

%
%
%
\section{Post-Widder's method of inverse Laplace transform}
\label{postwiddersmethod}
The inverse Laplace transform is usually calculated using a Bromwich contour integral over the complex plane. 
An alternative method is to introduce a method based on real variables, which is the Post-Widder's method. 
In this section we introduce the generalization of Post-Widder's method for the inverse of $q$-Laplace 
transfrom for both the single variable and the multivariable case.  

\subsection{Single variable inverse Laplace transform:}  
For the single variable inverse $q$-Laplace transform to derive the $q$-Widder's formula we will have to 
rewrite the type I kernel as $K_{q}(s,t) = \exp\left( \frac{1}{1-q}  \ln (1-(1-q)st) \right)$.  We then have to take 
the $k^{th}$ derivative of the Laplace transform and scale the function $f(t)$ to $f(\xi_{m} t)$ and the 
resulting expression is 
\begin{equation}
  F_{q}^{(k)} (s) = \int_{0}^{\infty}  dt  [-(1-q)t]^{k}   \frac{\Gamma \left( \frac{2-q}{1-q}   \right)}{\Gamma \left( \frac{2-q}{1-q} - k \right)}
                           (1-(1-q)st)^{\frac{1}{1-q} - k}    f(t \xi_{m}). 
\end{equation}
Through a change of variables $t=xy$ and $s=k/x$ we get: 
\begin{equation}
 F_{q}^{(k)} \left(  \frac{k}{x}  \right) = [-(1-q)]^{k} x^{k+1}   \frac{\Gamma \left( \frac{2-q}{1-q}  \right)}{\Gamma \left( \frac{2-q}{1-q} - k \right)}
    \int_{0}^{\infty} (1-(1-q) ky)^{\frac{1}{1-q} - k }  y^{k} dy  f(x y \xi_{m}). 
\end{equation}
The function $y^{k} (1-(1-q)k y)^{\frac{1}{1-q} - k}$ has a single maximum which is sharply peaked at $y=1$.   Hence we can replace 
the function $f(\xi_{m} xy)$ by $f(\xi_{m}x)$ and get
\begin{equation}
 F_{q}^{(k)} \left(  \frac{k}{x}  \right) = f(\xi_{m}x) [-(1-q)]^{k} x^{k+1}   \frac{\Gamma \left( \frac{2-q}{1-q} \right)}
           {\Gamma \left( \frac{2-q}{1-q} - k \right)} \int_{0}^{\infty} (1-(1-q) ky)^{\frac{1}{1-q} - k }  y^{k} dy. 
\label{kderiveqn}           
\end{equation}
The solution of the integral in the above equation is 
\begin{equation}
 \int_{0}^{\infty} (1-(1-q) ky)^{\frac{1}{1-q} - k }  y^{k} dy = \frac{\Gamma(k+1)   \Gamma \left(\frac{2-q}{1-q} -k \right)}
   {\Gamma \left( \frac{3-2q}{1-q} \right)  (1-q)^{k+1} k^{k+1}   }.
\end{equation}
Replacing the value of the integral  in Eq. (\ref{kderiveqn}) and simplifying we have 
\begin{equation}
F_{q}^{(k)} \left( \frac{k}{x} \right)  = (-1)^{k}  f(x \xi_{m}) x^{k+1}  \frac{\Gamma(k+1)}{k^{k+1} (2-q) }.
\end{equation}
where $\xi_{m} = \left[ \frac{1+(1-q)}{Q_{m}(2-q)} \right]^{\frac{1}{m-1}}$ valid for $m \geq 2$ and ill defined for $m=1$ and the polynomial
$Q_{m}(q) = \prod_{j=1}^{m} (1-(1-q)j)$ exists for only for $j \geq 1$. Substituting $t=\xi_{m} x$, the $q$-deformed 
Widder's formula for the inverse Laplace transform is 
\begin{equation}
f(t) =  \lim_{k \rightarrow \infty}  \frac{(-1)^{k}} {\Gamma(k+1)}  F_{q}^{(k)} (s) (2-q) s^{k+1}  |_{s=\frac{k \xi_{m}} {t}}.
\label{qwiddersfinalformula}
\end{equation}

\subsection{Multivariable inverse Laplace transform:}  
The multivariable inverse $q$-Laplace transform based on real variables  i.e., the Post-Widder's method for the type-I kernel is 
\begin{eqnarray}	
f(t_{1},t_{2},\ldots,t_{n}) &=& Q_{1}(2-q) \left(\prod_{l=1}^{n} \displaystyle \lim_{k_{l} \rightarrow \infty} \frac{(-1)^{k_{l}}}{\Gamma(k_{l} +1)} s_{l}^{k_{l} +1} \right)  \nonumber \\
&  & \frac{d^{k_{1}}}{d s_{1}^{k_{1}}} \left( \frac{d^{k_{2}}}{d s_{2}^{k_{2}}}\left( \ldots \frac{d^{k_{n}}}{d s_{n}^{k_{n}}} 
        F_{q} (s_{1},s_{2},\ldots,s_{n})\Biggr|_{s_{n}}  \ldots \right) \Biggr|_{s_{2}} \right) \Biggr|_{s_{1}}.
\end{eqnarray}
Here $s_{1} = (k_{1}/t_{1}) \xi_{r}$, where $r = \sum_{\ell=1}^{n} m_{\ell} + n$, $s_{2} = (k_{2}/t_{2}) (t_{1} s_{1} / k_{1})^{\frac{m_{2} +1}{m_{2}}}$ 
and  $s_{n} = (k_{n}/t_{n}) (t_{1} s_{1} / k_{1})^{\frac{m_{n} +1}{m_{n}}}$.   The factor $\xi$ is as defined in the previous section on single variable inverse $q$ 
Laplace transform based on Widder's method.

%
%
%
\section{Inverse Laplace transform for some elementary functions}
In this section we evaluate the Laplace transform of a simple algebraic function and calculate its inverse. 

\subsection{Laplace transform of single variable}
Let us consider an algebraic function $f(t)=t^{m-1}$, the $q$-Laplace transform of this 
function is 
\begin{equation}
L_{q}\left[t^{m-1}\right] = \displaystyle \int_{0}^{\infty} dt \exp_{q}(-st) t^{m-1}.
\label{qlap_alg}
\end{equation}
Evaluating the Laplace transform we get
\begin{equation}
F_{q}(s) = \frac{\Gamma(m)}{Q_{m}(2-q)} \frac{1}{s^{m}}\;, \qquad \mbox{for}\; m \geq 2.
\end{equation}
The inverse of the $q$-Laplace transform can be computed using a $q$-version of the Widder's formula. 
For this we calculate the $k^{th}$-derivative of $F_{q}(s)$
\begin{equation}
F_{q}^{(k)}(s) = \frac{1}{Q_{m}(2-q)} (-1)^{k} \frac{\Gamma(m+k)}{s^{m+k}}.
\label{algkderiv}
\end{equation}
Substituting Eq. (\ref{algkderiv}) in the $q$-Widder's formula we get
\begin{eqnarray}
 f(t) &=& \displaystyle \lim_{k \rightarrow \infty} \frac{(-1)^{k}}{\Gamma(k+1)}
		 \left( \frac{1}{Q_{m}(2-q)} (-1)^{k} \frac{\Gamma(m+k)}{s^{m+k}}\right) 
		 Q_{1}(2-q) s^{k+1}\Biggr|_{s = \frac{k \xi_{m}}{t}}, \nonumber \\
	 &=& \displaystyle \lim_{k \rightarrow \infty} \frac{(-1)^{k}}{\Gamma(k+1)}
		 \left( \frac{1}{Q_{m}(2-q)} (-1)^{k} \frac{\Gamma(m+k)}{s^{m-1}}\right) 
		 Q_{1}(2-q) \Biggr|_{s = \frac{k \xi_{m}}{t}}.
\end{eqnarray} 
where, $\xi_{m} = \left(\frac{Q_{1}(2-q)}{Q_{m}(2-q)}\right)^{\frac{1}{m-1}}$ and the subscript `$m$' depends 
on the exponent of `$s$', since the exponent of `$s$' is `$m-1$' the value of `$m$' is `$m$'.  Substituting the 
value of `$s$' we get:
\begin{eqnarray}
f(t) &=& \frac{Q_{1}(2-q)}{Q_{m}(2-q)} \frac{t^{m-1}}{\xi^{m-1}} 
		 \left(\displaystyle \lim_{k \rightarrow \infty} \frac{ k^{1- m}  \Gamma(m+k)}{\Gamma(k+1)} \right). 
		 \label{q_ilap_ag_wid}
\end{eqnarray}
On substitution of the limits and simplifying Eq.  (\ref{q_ilap_ag_wid}) we get the algebraic limit $f(t) = t^{m-1}$.
This  validates the Post-Widder's method of computing inverse Laplace transform for any Algebraic function.  
The Laplace transform and inverse Laplace transform of some common functions is given in Table \ref{table1} below.
Here we have calculated the inverse transform using the contour integration method as well as the Post-Widder's 
technique.

\begin{table}
\caption{Laplace and inverse Laplace transforms of some elementary functions }
\begin{tabular}{|c|c|c|c|c|}
\hline
S.No. & $f(t)$ & $F_q(s)$      \\
\hline
1&   $t^{m-1} $   & $\frac{\Gamma(m)}{Q_m(2-q)}\frac{1}{s^{m}}$\\
\hline
2 &  $\exp(\pm \sigma_{\epsilon} t)  = \left\{ \begin{array}{ll} 
                   \exp(\pm\alpha t) & \mbox{for} \;\epsilon=0   \\ 
                   \exp(\pm i\alpha t) & \mbox{for} \; \epsilon=1 
                   \end{array} \right. $
& $\frac{1}{sQ_1(2-q)}{}_1F_{1}\left(1;a;\pm\frac{\sigma_\epsilon }{(1-q)s}\right)$  \\
\hline
3 & $\exp(-\alpha t^{2})$ & $\frac{1}{sQ_1(2-q)}{}_2F_{2}\left(1,\frac{1}{2};a/2,b/2;-\frac{\alpha}{(1-q)^{2}s^{2}}\right)$ \\
\hline
4 &  $\left\{ \begin{array}{ll}
		\cos\alpha t& \mbox{for} \;\delta=0\\
		\sin\alpha t& \mbox{for} \;\delta=1 \end{array}\right.$   &  $\frac{\alpha^{\delta}}{s^{\delta+1}}   \frac{1}{Q_{\delta +1}(2-q) } {}_1F_{2}\left(1; \frac{a +\delta}{2},\frac{b+\delta}{2}; 
		-\frac{\alpha^{2}}{4(1-q^{2})s^{2}}\right) $ \\
\hline
5 & $\left\{ \begin{array}{ll}
		\cosh \alpha t& \mbox{for} \;\delta=0\\
		\sinh \alpha t& \mbox{for} \;\delta=1 \end{array}\right.$  & $\frac{\alpha^{\delta}}{s^{\delta+1}}   \frac{1}{Q_{\delta +1}(2-q) } {}_1F_{2}\left(1; \frac{a +\delta}{2},\frac{b+\delta}{2}; 
		\frac{\alpha^{2}}{4(1-q^{2})s^{2}}\right) $ \\
\hline
\multicolumn{3}{|c|}{$q$ - Generalized functions} \\
\hline
1 & $\exp_{q^{\prime}}(\pm\sigma_\epsilon t)=\left\{\begin{array}{ll}
		\exp_{q^{\prime}}(\pm\alpha t)& \mbox{for} \;\epsilon=0\\
		\exp_{q^{\prime}}(\pm i\alpha t)& \mbox{for} \;\epsilon=1\end{array}\right.$  &  
		$\frac{1}{sQ_1(2-q)}{}_2F_{1}\left(1, c ; a ;\mp\frac{(1-q^{\prime})\sigma_\epsilon}{(1-q)s}\right) $    \\
\hline
2 & $\exp_{q^{\prime}}(-\alpha t^{2})$ & $\frac{1}{sQ_1(2-q)}{}_3F_{2}\left(1,\frac{1}{2},c;a/2,b/2;\frac{(1-q^{\prime})\alpha}{(1-q)^{2}s^{2}}\right)$\\
\hline
3 & $\left\{ \begin{array}{ll}
		\cos_{q^{\prime}} \alpha t & \mbox{for} \;\delta=0\\
		\sin_{q^{\prime}} \alpha t & \mbox{for} \;\delta=1 \end{array}\right.$ &    
         $ \frac{\alpha^{\delta}}{s^{\delta+1} Q_{\delta +1}(2-q) } {}_3F_{2}\left(1, \frac{c+\delta}{2}, \frac{c-1+\delta}{2};  \frac{a +\delta}{2},\frac{b+\delta}{2}; 
		-\frac{ (1-q^{\prime})^{2}  \alpha^{2}}{4(1-q)^{2}s^{2}}\right) $ \\
\hline
4 & $\left\{ \begin{array}{ll}
		\cosh_{q^{\prime}} \alpha t & \mbox{for} \;\delta=0\\
		\sinh_{q^{\prime}} \alpha t & \mbox{for} \;\delta=1 \end{array}\right.$ &  $\frac{\alpha^{\delta}}{s^{\delta+1} Q_{\delta +1}(2-q) } 
		{}_3F_{2}\left(1, \frac{c+\delta}{2}, \frac{c-1+\delta}{2};  \frac{a +\delta}{2},\frac{b+\delta}{2}; 
		\frac{ (1-q^{\prime})^{2}  \alpha^{2}}{4(1-q)^{2}s^{2}}\right) $ \\
\hline		
\multicolumn{3}{|c|}{In the table above we use the following definitions  \qquad  $a = \frac{3-2q}{1-q}$, \quad $b = \frac{4-3q}{1-q}$,  \quad $ c = - \frac{1}{1-q^{\prime}}$ } \\
\hline
\end{tabular} 
\label{table1}
\end{table}

\subsection{Multivariable Laplace transform:}
To illustrate the multivariate Laplace trasnform let us choose the function 
$f(t_{1},t_{2},\ldots,t_{n}) = \displaystyle \alpha\; \prod_{l=1}^{n} t_{l}^{m_{l}}$ where $\alpha$ is a constant. 
The multivariate $q$-Laplace transform of this function is 
\begin{eqnarray}
F_{q} (s_{1},s_{2},\ldots,s_{n}) = \alpha \frac{\Gamma\left(\frac{1}{1-q} + 1\right)}{\Gamma\left( \frac{1}{1-q} + \displaystyle \sum_{l=1}^{n} m_{l} + n + 1\right)} 
					    \times \displaystyle \prod_{l=1}^{n} \; \frac{\Gamma\left(m_{l} + 1\right)}{\left(1-q\right)^{m_{l}+1} s_{l}^{m_{l}+1}} 
\label{Fq}
\end{eqnarray}
To calculate the inverse we can substitute $F_{q} (s_{1},s_{2},\ldots,s_{n})$ in the expression for general inverse 
\begin{eqnarray}
	f(t_{1},t_{2},\ldots,t_{n}) &=& \alpha \frac{Q_{n}(2-q)}{(2\pi i)^{n}}  \frac{\Gamma\left(\frac{1}{1-q} + 1\right)}{\Gamma\left( \frac{1}{1-q} + \displaystyle \sum_{l=1}^{n} m_{l} + n + 1\right)} \displaystyle \prod_{l=1}^{n} \; \frac{\Gamma\left(m_{l} + 1\right)}{\left(1-q\right)^{m_{l}+1}} \times \nonumber \\
	& &    \int_{\Gamma_{1}} ds_{1} \int_{\Gamma_{2}} ds_{2} \ldots \int_{\Gamma_{n}} ds_{n}  
	\left[1 - (1 - q) \left(\displaystyle \sum_{l=1}^{n} s_{l} t_{l} \right) \right]^{\frac{(n+1)q - (n+2)}{1-q}} 
	\frac{1}{s_{l}^{m_{l}+1}} \nonumber \\
	&=&  \alpha \frac{Q_{n}(2-q)}{(2\pi i)^{n} Q_{\left(\displaystyle \sum_{j=1}^{n} m_{j} + n\right)}(2-q)} \displaystyle \prod_{l=1}^{n} \; \Gamma\left(m_{l} + 1\right)    \\
	& & \int_{\Gamma_{1}} ds_{1} \int_{\Gamma_{2}} ds_{2} \ldots \int_{\Gamma_{n}} ds_{n}  
	\left[1 - (1 - q) \left(\displaystyle \sum_{l=1}^{n} s_{l} t_{l} \right) \right]^{\frac{(n+1)q - (n+2)}{1-q}} 
	\frac{1}{s_{l}^{m_{l}+1}} \nonumber 
\end{eqnarray}
Using complex contour integration we get $f(t_{1},t_{2},\ldots,t_{n}) = \displaystyle \alpha\; \prod_{l=1}^{n} t_{l}^{m_{l}}$, where $\alpha$ is a constant.
Alternatively we can use the Post-Widder's method to get 
\begin{eqnarray}	
f(t_{1},t_{2},\ldots,t_{n}) &=& Q_{1}(2-q) \left(\prod_{l=1}^{n} \displaystyle \lim_{k_{l} \rightarrow \infty} \frac{(-1)^{k_{l}}}{\Gamma(k_{l} +1)} s_{l}^{k_{l} +1} \right) \nonumber  \\
& &  \frac{d^{k_{1}}}{d s_{1}^{k_{1}}} \left( \frac{d^{k_{2}}}{d s_{2}^{k_{2}}}\left( \ldots \frac{d^{k_{n}}}{d s_{n}^{k_{n}}} F_{q} (s_{1},s_{2},\ldots,s_{n})\Biggr|_{s_{n}}  \ldots \right) \Biggr|_{s_{2}} \right) \Biggr|_{s_{1}}
\label{mvltalgebraic}
\end{eqnarray}
Substituting $F_{q} (s_{1},s_{2},\ldots,s_{n})$ in Eq. (\ref{mvltalgebraic}), we get $f(t_{1},t_{2},\ldots,t_{n}) = \displaystyle \alpha\; \prod_{l=1}^{n} t_{l}^{m_{l}}$.   Thus we 
evaluate the inverse Laplace transform of the multivariable algebraic function using complex contour integration method as well real variable based Post-Widder's method. 
We find identical results which leads to conclude that both these methods are equivalent.

\section{Concluding Remarks}
\label{conclusion}
A generalization of the Laplace transform based on Tsallis $q$-exponential is investigated in the present work.  We use 
$K_{I} (q;s,t) = \exp_{q}(-st)$ as the kernel of the transform.  A single variable Laplace transform has been defined in 
 \cite{naik2016q} based on this Kernel.  But the inverse transformation has not been defined so far.  In the present 
 work we define the inverse Laplace transform of a single variable function using the complex integration method as well as 
 the Post-Widder's method which uses real variables.  Then we introduce a multivariable Laplace transform for the same 
 kernel and also define the inverse transform through the complex integration method as the real variable based method. 
 We verify the properties of the Laplace transform as well as the inverse transform and compute them for some 
 elementary functions.  The results are given in Table: \ref{table1}, where the inverse transforms were calculated using 
 the complex integration method and the real variable based Post-Widder's method.  

An application of the Laplace transform and its inverse in statistical mechanics is to interrelate the partition function and 
the density of states.  To describe a physical system in thermodynamic equilibrium with its surroundings we need to 
describe its thermal, mechanical and chemical properties.  Each of this property is characterized using a pair of 
quantities of which one is is an extensive variable and the other is an intensive variable.  So, in total we have $8$ 
 $(2^{3})$ different ensembles.  Of the eight ensembles there are four of them where the temperature is fixed 
 and these are known as the isothermal ensembles and the remaining four where the heat is fixed are the adiabatic 
 ensembles.  The Laplace transform and the inverse transform can be used to interrelate the partition functions and 
 density of states of the different ensembles.  For example let us consider a classical ideal gas in $D$-dimensions
 with the Hamiltonian $H= \sum_{i=1}^{DN} \frac{p_{i}^{2}}{2m}$ and the partition of this system is 
\begin{equation}
 Z_{q}(\beta) = \frac{V^{N}(2\pi m)^{\frac{DN}{2}}}{h^{DN} N!} 
	           \left(\frac{\Gamma\left(\frac{1}{1-q} +1 \right)}{(1-q)^{\frac{DN}{2}}\; 
	           \Gamma\left(\frac{1}{1-q} + \frac{DN}{2} + 1\right)}\right) \frac{1}{\beta^{\frac{DN}{2}}}.
\label{pfidealgas}	           
\end{equation} 
From the generalized Post-Widder's method we get: 
\begin{equation}
g(E) = \displaystyle \lim_{k \rightarrow \infty} \frac{(-1)^{k}}{\Gamma(k+1)} 
         Z_{q}^{(k)}(\beta) \beta^{k+1} Q_{1}(2-q) \Biggr|_{\beta = \frac{k \xi_{m}}{E}}
        =  \frac{V^{N}(2\pi m)^{\frac{DN}{2}}}{h^{DN} N!\left(\frac{DN}{2}-1\right)!} E^{\frac{DN}{2}-1}. 
\label{dosidealgas}         
\end{equation}
which is the density of states of the classical ideal gas.   Similarly  we can consider a collection of $N$ harmonic 
oscillators in $D$-dimensions with the Hamiltonian 
$H= \sum_{i=1}^{DN} \left(\frac{p_{i}^{2}}{2m} + \frac{1}{2} m \omega^{2} x_{i}^{2} \right)$.  The partition function 
of this system is 
\begin{equation}
Z_{q}(\beta) = \frac{1}{\left(\hbar\omega\right)^{DN}} 
	\left(\frac{\Gamma\left(\frac{1}{1-q} +1 \right)}{(1-q)^{DN}\; \Gamma\left(\frac{1}{1-q} + DN + 1\right)}\right) \frac{1}{\beta^{DN}}.
\label{partitionfunctionoscillator}
\end{equation} 
Using the Post-Widder's method of inverse transform we get 
\begin{eqnarray}
g(E) = \displaystyle \lim_{k \rightarrow \infty} \frac{(-1)^{k}}{\Gamma(k+1)} Z_{q}^{(k)}(\beta) \beta^{k+1} Q_{1}(2-q) 
         \Biggr|_{\beta = \frac{k \xi}{E}} 
         =  \frac{1}{\left(\hbar \omega\right)^{DN}\left(DN-1\right)!} E^{DN-1} 
\label{densityofstates}
\end{eqnarray}
which is the expression for the density of states of the system in the microcanonical ensemble.  While these applications 
demonstrate the generalized Laplace transform and its inverse, we can also use them in the fields of 
atomic physics, open quantum systems and also in finding the solutions of differential equations.  

\section*{Acknowledgements}
Md.~Manirul Ali was supported by the Centre for Quantum Science and Technology, Chennai Institute of Technology, India,
vide funding number  CIT/CQST/2021/RD-007.
R. Chandrashekar was supported in part by a seed grant from IIT Madras to the Centre for
Quantum Information, Communication and Computing.

\section*{Author Declarations}
\subsection*{Conflict of interest}  
The authors have no conflicts to disclose

\subsection*{Data availability}
Data sharing is not applicable to this article as no new data were created or analyzed in this study.

\bibliography{reference}

\end{document}